\pdfoutput=1 
\documentclass[%
reprint,
superscriptaddress,
amsmath,amssymb,
aps,
floatfix,
]{revtex4-2}
\usepackage{import}
\usepackage{xcolor}
\usepackage{natbib}
\usepackage{floatrow}
\usepackage{graphicx}
\usepackage{dcolumn}
\usepackage{bm}
\usepackage{lipsum}
\usepackage[normalem]{ulem}

\usepackage{float}
\usepackage{pifont}
\usepackage{hyperref}
\usepackage{changes}
\usepackage{comment}

\begin{document}
	
	\preprint{APS/123-QED}
	
	\title{Microwave-to-optical conversion in a room-temperature $^{87}$Rb vapor for frequency-division multiplexing control
	}
	
	\author{Benjamin D. Smith}
	\email{bdsmith@ualberta.ca}
	\affiliation{%
		Department of Physics, University of Alberta, Edmonton, Alberta T6G 2E1, Canada\\
	}
	
	\author{Bahar Babaei}%
	\affiliation{%
		Department of Physics, University of Alberta, Edmonton, Alberta T6G 2E1, Canada\\
	}

	\author{Andal Narayanan}%
	\affiliation{%
		Department of Physics, University of Alberta, Edmonton, Alberta T6G 2E1, Canada\\
	}
	\affiliation{Raman Research Institute, Bangalore-560080, India.}

	\author{Lindsay J. LeBlanc}%
	\email{lindsay.leblanc@ualberta.ca}
	\affiliation{%
		Department of Physics, University of Alberta, Edmonton, Alberta T6G 2E1, Canada\\
	}

	\date{\today}
	
	\begin{abstract}
		\hspace{5.9 cm }{\color{black} \large Abstract}\\
		Coherent microwave-to-optical conversion is crucial for transferring quantum information generated in the microwave domain to optical frequencies, where propagation losses can be minimized. Coherent, atom-based transducers have shown rapid progress in recent years. This paper reports an experimental demonstration of coherent microwave-to-optical conversion that maps a microwave signal to a large, tunable 550(30) MHz range of optical frequencies using room-temperature $^{87}$Rb atoms. The inhomogeneous Doppler broadening of the atomic vapor advantageously supports the tunability of an input microwave channel to any optical frequency channel within the Doppler width, along with the simultaneous conversion of a multi-channel input microwave field to corresponding optical channels. In addition, we demonstrate phase-correlated amplitude control of select channels, providing an analog to a frequency domain beam splitter across five orders of magnitude in frequency. With these capabilities, neutral atomic systems may also be effective quantum processors for quantum information encoded in frequency-bin qubits.
	\end{abstract}

	\maketitle
	
	\section{Introduction}
	\begin{figure*}
		\includegraphics{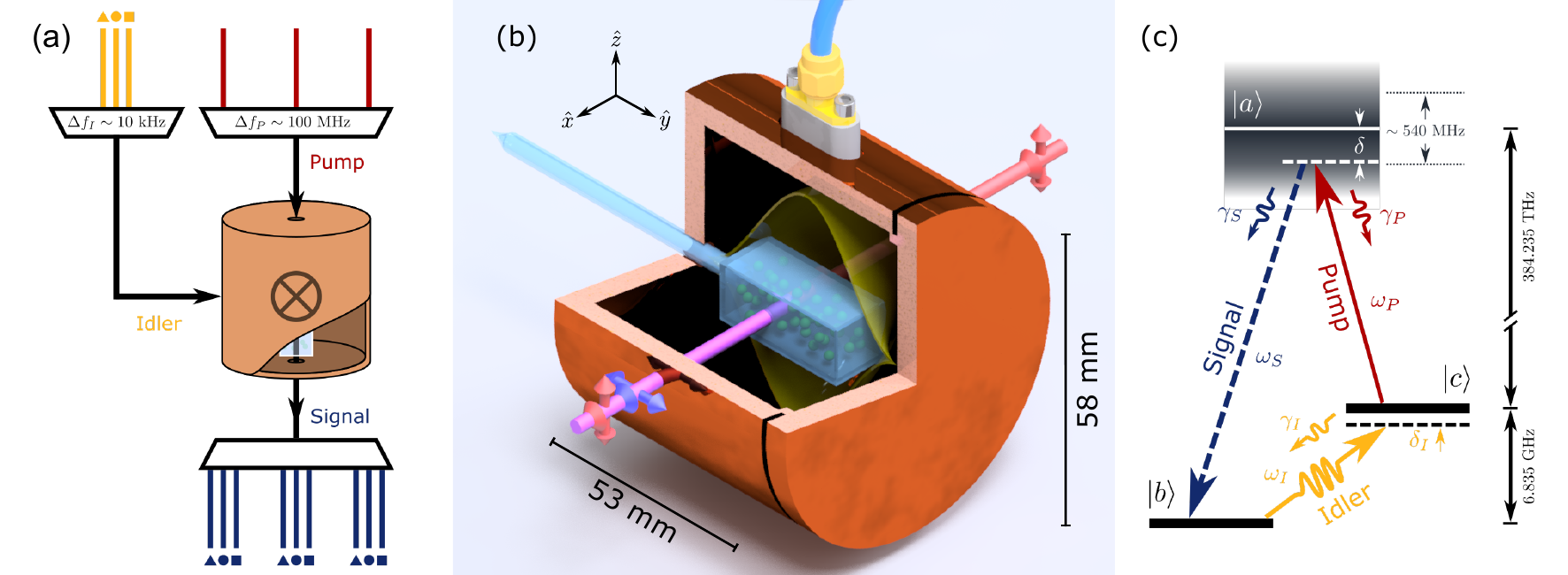}
		\caption{ {\color{black} \bf{ Main components of the experiment.}} (a) Schematic of the atomic frequency-division multiplexing scheme. The microwave cavity-vapor cell system acts analogously to an RF mixer, and information encoded in idler {\color{black} (yellow)} channels (\ding{115}, \ding{108}, \ding{110}) is combined coherently with multiple pump {\color{black} (red)} channels to produce many output signal {\color{black} (blue)} channels over a range of optical frequencies. (b) A representation of our microwave cavity and the generation process. The optical pump beam passes transversely through the cavity and cell and is polarized along the $z$-axis, parallel to a weak magnetic field.
			The cavity's $\rm TE_{011}$ mode overlaps with the pump interaction region inside the vapor cell. The arrows define the beam propagation and polarization directions. (c) Three-level energy diagram showing the cyclical transition scheme. The idler ($\omega_I$) and pump ($\omega_P$) fields combine coherently in the vapor to produce the signal field ($\omega_S$). The frequencies shown are for $^{87}$Rb; in a room-temperature vapor, the excited state addressed by the pump field is Doppler-broadened by approximately 550 MHz.}
		\label{fig:intro} 
	\end{figure*}
	
	Quantum information processing using solid-state qubits, such as superconducting qubits or spins associated with defects in solid state systems, has shown {\color{black} rapid} progress in the field of quantum information technology \cite{Arute2019, Son2020}. The typical operation frequency of these devices is $\sim$10 GHz, in the microwave region of the electromagnetic spectrum. Due to the inherent room-temperature thermal-photon occupation of cables at GHz frequencies that leads to signal noise, practical implementation of scalable and distributed quantum networks at these frequencies poses a formidable challenge~\cite{Storz2023}. In recent years, progress has been made in addressing coherent conversion of signals from the microwave to optical regimes \cite{Lambert2020, Lauk2020}: at optical frequencies, not only are the thermal noise photons negligible at room temperature, but there exist efficient single-photon detectors and technologies for quantum state storage and reconstruction.\\
	
	\indent Several systems have been used to coherently convert signals at microwave frequencies to the  optical regime. These include electro- and magneto-optical devices \cite{Fan2018, Hisatomi2016}, opto-mechanical structures, \cite{Higginbotham2018} and atoms with suitable internal energy level spacings~\cite{Han2018, Adwaith2019, Tu2022, Borowka2023}. Impressive progress on improving the efficiency of the conversion process led to efficiency near 80\%~\cite{Tu2022}, and the largest reported conversion bandwidths of an input microwave signal are 15-16 MHz~\cite{Vogt2019, Zhu2020, Borowka2023}. Along with conversion bandwidth, a transducer that can accommodate several channels of input and output frequencies is advantageous, and tunability over input microwave frequencies across 3 GHz width was recently reported~\cite{Shen2022}. Yet, transducers that operate over multiple optical output frequencies have not been demonstrated to date. \\
	
	\indent In this article we show that a limited bandwidth input microwave signal can be converted to a large tunable range of output optical frequencies, across 550(30) MHz. This capacity enables information encoded in a narrow-bandwidth microwave field to be mapped to several frequencies in the optical domain, achieving coherent frequency-division multiplexing (FDM). The ability to tune the optical signal output over such a large frequency range arises because of the presence of a large inhomogeneous optical Doppler width for Rb atoms at room temperature. The Doppler width also enables an input multiple-channel  microwave idler to be simultaneously and coherently converted to a multiple-channel optical-signal output. In addition, we demonstrate correlated amplitude control of the converted multi-channel optical field with the ability to selectively extinguish a desired output frequency channel.  This action is very similar to the frequency domain beam-splitting operation for frequency bins, demonstrated using electro-optic modulators (EOMs) and pulse shapers \cite{Lu2018a}.\\
	
	\section{Results}
	
	Coherent frequency-division multiplexing enables qubits encoded in a microwave frequency to be placed in any desired output optical channel within the range of tunability. Multiple qubits in different input microwave channels can also be coherently mapped to the optical domain. The ability to control the amplitude of selected output optical channels is very useful in performing single-qubit-gate operations spanning involving both microwave and optical frequencies. With a channel linewidth of about 30 Hz, this results in a large channel capacity of about $10^{7}$  channels over the output tunable range. Thus neutral atom systems are a promising candidate as a quantum frequency processor platform \cite{Lukens2019} for quantum information encoded in frequency bin qubits.

	\begin{figure}[tbh!p]
		\includegraphics{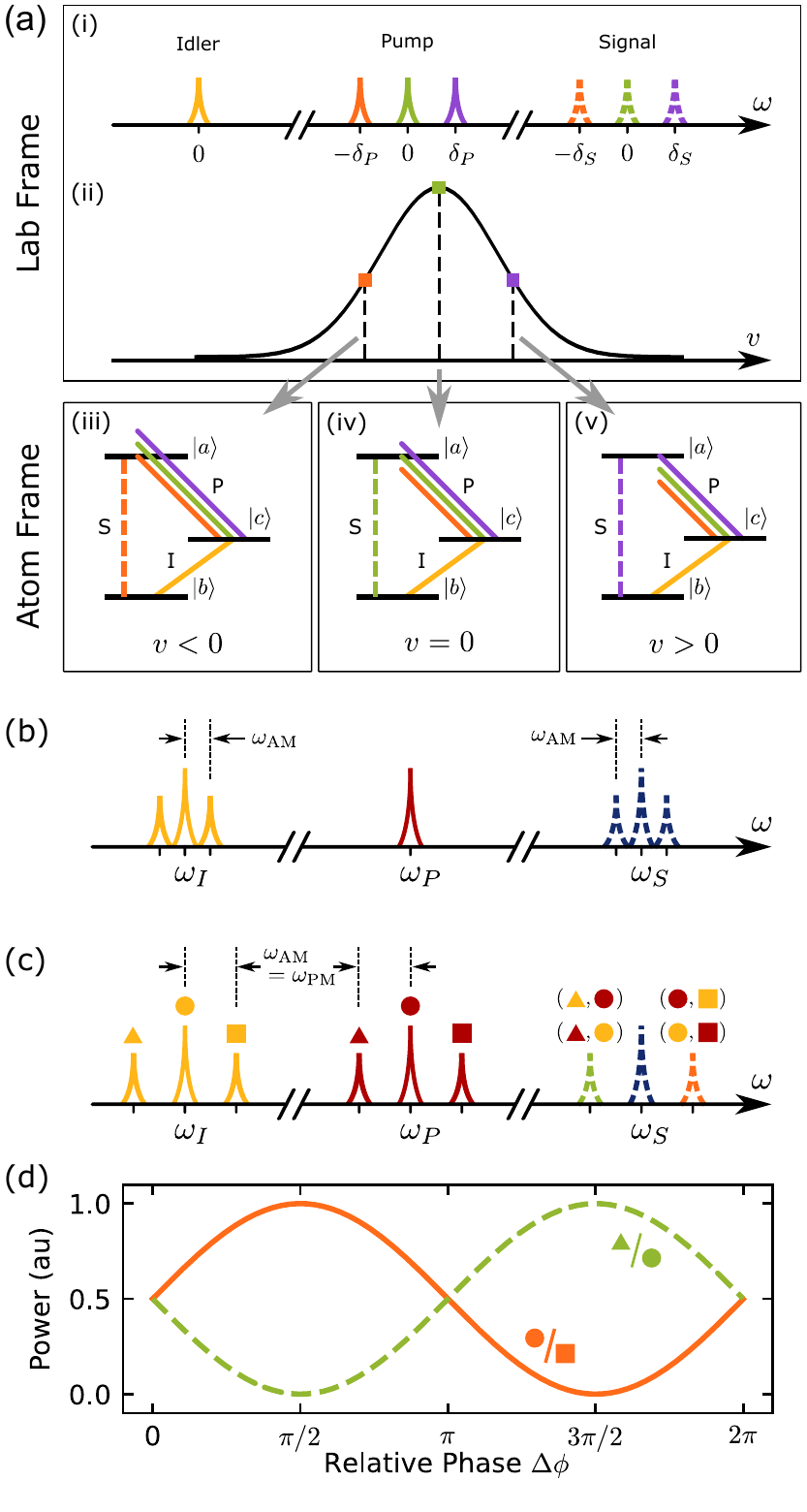}
		\caption{ {\color{black} \bf{Numerical simulations:}}(a-i) Lab-frame cartoon depicting multiple pump frequency channels {\color{black}(central-green, upper-purple, lower-organge)} combining with a single idler {\color{black}(yellow)} to produce multiple signal channels. {\color{black}(a-ii)} These pump frequencies access independent atomic velocity classes from a thermal MB distribution. {\color{black}(a-iii to a-v)} In their reference frame, atoms in these velocity classes see the respective pump frequency Doppler-shifted to resonance. Generation occurs in the velocity class for which the three-photon resonance condition (i.e. energy conservation) is satisfied. (b) Calculated spectra of an amplitude-modulated idler field showing sidebands (SBs); the carrier and SBs combine with the pump to produce a multichannel signal spectrum. (c) When the idler and pump are both modulated at the same frequency, multiple generation pathways interfere producing an upper and lower generated signal SBs. {\color{black} The input fields each have lower (\ding{115}), central (\ding{108}), and upper (\ding{110}) spectral peaks.} (d) Calculated curves showing correlated signal SB amplitudes, which vary coherently with the relative phase of the two modulations depicted in (c). Both AM and PM modulation indices are 0.2.}
		\label{fig:theory} 
	\end{figure}
	\subsection{Numerical model}
	Microwave-to-optical conversion in a room-temperature $^{87}$Rb atomic vapor is possible through a sum-frequency generation (SFG) process analogous to a RF mixer [Fig.~\ref{fig:intro}(a)] and depicted in Fig.~\ref{fig:intro}(c). The nonlinear process uses a hybrid, magneto-optical second-order $[\chi^{(2)}]$ susceptibility induced in the vapor by the interaction of three energy levels with an optical pump (electric-dipole transition) and a microwave idler (magnetic-dipole transition) fields. (Note that isotropic atomic systems with only electric-dipole transitions have a $\chi^{(2)}$ of zero.) The idler field ($\omega_{\text{I}}$) is supported by the $\rm TE_{011}$ mode of a cylindrical copper microwave cavity~\cite{Tretiakov2020, Ruether2021} [Fig.~\ref{fig:intro}(b)] and connects the $|b\rangle \equiv |5S_{1/2}, F = 1 \rangle$ and $|c\rangle \equiv |5S_{1/2}, F = 2 \rangle$ ground state levels. The pump frequency $\omega_{\text{P}}$ is, chosen anywhere within the Doppler width {\color{black} [Fig.~\ref{fig:intro}(c)]}, and connects the $|c\rangle$ and  $|a\rangle \equiv | 5P_{3/2}, F^{\prime} = 1, 2 \rangle$ levels. We define $\delta$ as the detuning of the pump field from the the mid-energy point between the $|F^{\prime}=1\rangle$ and $|F^{\prime}=2\rangle$ excited levels. In SFG, the pump combines coherently with the idler to generate a new optical signal field at frequency $\omega_{\text{S}}$ = $\omega_{\text{P}} + \omega_{\text{I}}$.

	We begin by establishing a numerical model that explains the large frequency tunability and multi-channel conversion capability observed in this experiment, similar to that discussed for a generic, three-level, room-temperature atomic system, with interacting idler, pump, and signal fields \cite{Adwaith2021}. The Rabi frequencies of the idler and pump fields at position $z$ are $\Omega_{\text{I}}({z}) = \boldsymbol{\mu}_{\text{cb}} \cdot \mathbf{B}_{\rm I}(z)$ and 
	$\Omega_{\text{P}}({z}) = \mathbf{d}_{\text{ac}} \cdot \mathbf{E}_{\rm P}(z)$, where $\Delta \varphi$ is the relative phase between the fields, and $\boldsymbol{\mu}_{\text{cb}}$ and $\mathbf{d}_{\text{ac}}$ are the dipole matrix elements between $|c \rangle~\&~|b \rangle$ and $|a \rangle~\&~| c \rangle$ levels. The signal field is generated due to a sum-frequency interaction between the idler and pump fields. The Rabi frequency of the generated signal field $\Omega_{\text{S}}({\ell})$ after a medium of length $\ell$, under a rotating wave and three-photon transformation, is (see {\color{black} Supplementary Note 4})
	
	\begin{equation}
		\Omega_{\text{S}}({\ell}) = \frac{i \Omega_{\text{P}}(0) \Omega_{\text{I}}(0) e^{- i \Delta \varphi}}{2\Gamma_2} (e^{-\beta \ell} - 1),
		\label{eq:signal}
	\end{equation}
	where the constant
	\begin{equation}
		\beta~=~\frac{{2\eta_{\text{S}}\Gamma_2\Gamma_3}}{{4\Gamma_1\Gamma_2\Gamma_3 + \Gamma_3|\Omega_{\text{P}}(0)|^2 + \Gamma_2|\Omega_{\text{I}}(0)|^2}},
	\end{equation}
	
	and
	
	\begin{align}
		&\Gamma_1= \frac{\gamma_{\rm P}}{2} + \frac{\gamma_{\rm S}}{2} - i ~\Delta_\text{S} (\bm{v}), \nonumber \\
		&\Gamma_2=\frac{1}{2} \gamma_{\rm I}- i \left[\Delta_\text{S} (\bm{v}) - \Delta_\text{P} (\bm{v})\right], \label{eq:gamma2} \\
		&\Gamma_3=\frac{\gamma_{\rm P}}{2} + \frac{\gamma_{\rm S}}{2}+\frac{\gamma_{\rm I}}{2}- i~\Delta_\text{P} (\bm{v}), \nonumber \\
		&\eta_{\text{S}} (v)=~{\bm{d}_{\text{ac}}^2\omega_\textsubscript{s}N(v)}/{2 \epsilon_{0} c}, \nonumber
	\end{align}
	with $N(v)$ is the number density of atoms in the velocity class with $v = |{\bm v}|$. The value $N(v)$ is the Maxwell-Boltzmann (MB) velocity distribution of atoms at room temperature. The Doppler-shifted detunings of the pump and the generated signal fields, as seen from the atom frame [Fig. (\ref{fig:theory} a, iii-v)], are denoted as $\Delta_\text{P} =\delta_\text{P}- \bm{k}_\text{P}\cdot\bm{v}$ and $\Delta_\text{S}  =\delta_\text{S}- \bm{k}_\text{S}\cdot\bm{v}$, while the corresponding detunings in the lab frame are $\delta_\text{P}$ \& $\delta_\text{S}$ [Fig. \ref{fig:intro}(c) \& Fig. \ref{fig:theory}(a, i)]. The microwave idler field has negligible Doppler shift for all velocity values. The natural decay rates from the $|a\rangle$ level to both the $|c\rangle$ and $|b\rangle$ levels are denoted by $\gamma_{P}$ and $\gamma_{S}$, respectively [Fig. \ref{fig:intro} (c)].\\
	\indent To maximise the generated signal field, sufficiently high input pump and idler field intensities and a small  value for $\Gamma_2$ [Eq.~(\ref{eq:signal})] are required. From Eq.~(\ref{eq:gamma2}) this is achieved when 
	
	\begin{eqnarray}
		\Delta_\text{S} (\bm{v})-\Delta_\text{P} (\bm{v}) = 0.
		\label{eq:Doppler-eqn}
	\end{eqnarray}
	
	\indent Typically, the wavelengths of optical fields are such that $k \equiv |\mathbf{k}_S| \approx  |\mathbf{k}_P| \gg |\mathbf{k}_I|$, in which case Eq.~(\ref{eq:Doppler-eqn}) reduces to $(\delta_{\rm S} - \delta_{\rm P})  = 0$. Since the signal and pump fields travel in the same direction, the phase-matching condition $\mathbf{k}_{\rm I} + \mathbf{k}_{\rm P} = \mathbf{k}_{\rm S}$ is automatically satisfied for all atomic velocity classes.

	\indent The maximum signal generation for each velocity class is obtained when $\delta_{\text{S}}$ of the signal-field is the same as the pump-field detuning $\delta_{\text{P}}$, such that different velocity classes  participate in the maximum signal-field generation at different pump-field detunings. Therefore, changing the laboratory pump frequency within the Doppler width to $\omega_{\rm P}^\prime = \omega_{\rm P} + \delta$ gives rise to a corresponding signal field at $\omega_{\rm S}^\prime = \omega_{\rm S} + \delta$, thus enabling a large tunable generation bandwidth. This is depicted numerically in Fig \ref{fig:theory}(a, ii) for atoms at T = 300$^\circ$ C.  The shape of the generated signal field is primarily determined by the MB velocity distribution.
	
	\indent The broad Doppler width at room temperature and guaranteed phase-matching for sum-frequency generation [Eq.~(\ref{eq:Doppler-eqn})] make possible the simultaneous conversion of multi-channel microwave field inputs to a multi-channel optical signal outputs {\color{black} [Fig.~\ref{fig:theory}(b)]}. Multiple frequency channels in the idler and pump fields are obtained by either amplitude or phase modulation. Specifically, the idler field amplitude is modulated as  $\Omega_{\rm I}(t) = \Omega_{\rm I0}(1+ a_m\sin(\omega_{\rm AM}) t)$  and phase modulation of the pump field takes the form $\Omega_{\rm P}(t) = \Omega_{\rm P0}\sin(\omega_{\rm P} t + p_m \sin(\omega_{\rm PM})t + \Delta \phi)$. The symbols $a_m$ and $p_m$ denote amplitude and phase modulation indices. The modulation frequencies are $\omega_{\rm AM}$ and $\omega_{\rm PM}$ and $\Delta \phi$ denotes the relative phase between the two modulation fields. By multiplying the modulated idler and pump fields to obtain the signal field, we derive the amplitudes for the generated signal field and its first-order side-bands for the case of equal modulation frequencies $\omega_{\rm AM} = \omega_{\rm PM} = \delta$ and for equal modulation indices $a_p = a_m = 0.2$. At equal modulation frequencies, the pump and idler fields associated with the sideband  channels interfere with each other [Fig.~\ref{fig:theory}(c)]  enabling control of their amplitudes through $\Delta \phi$ \cite{Lu2018b}. The powers of the positive and negative sideband channels at $\omega_{\text{S}} \pm \delta$ are proportional to $a_m^2 + a_p^2 + 2 a_ma_p\sin(\Delta \phi)$
	and $a_m^2 + a_p^2 - 2 a_m a_p\sin(\Delta \phi)$ (see {\color{black} Supplementary Note 7}). The power variations at $\pm \delta$ sidebands, as a function of $\Delta \phi$, are shown in Fig.~\ref{fig:theory}(d).

	\begin{figure*}
		\centering
		\includegraphics{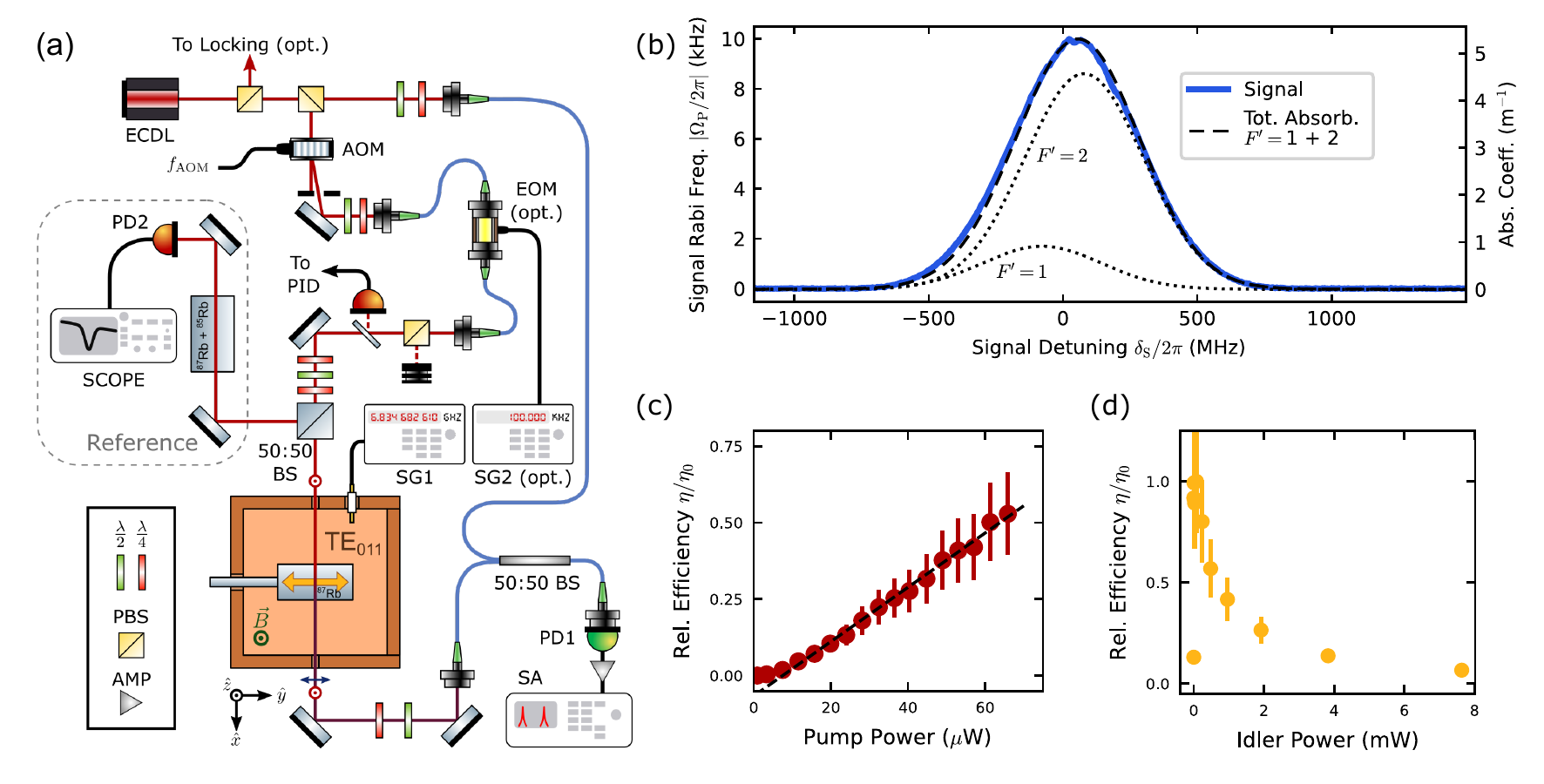}
		\caption{{\color{black} \bf{ Experimental setup and main results:}}(a) Schematic of the experimental setup: Central to the experiment is a microwave cavity supporting a $\rm TE_{011}$ mode, polarized along the $y$-axis. The light incident on the cavity is polarized along the $z$-axis and propagates along the $x$-axis. The vapor cell inside the cavity has enriched $^{87}$Rb. Unshifted light from the {\color{black}external cavity} diode laser (ECDL) is used as a local oscillator for heterodyne detection; it is combined with the other beam in a fiber 50:50 beam splitter (BS), producing two beat signals at $f_{\rm het,~P}$ and $f_{\rm het,~S}$. AOM - Acousto-optic modulator; EOM - Electro-optic modulator; SG - signal generator; SA - spectrum analyzer; PD - photodiode. 
			(b) The generated signal measured with the spectrum analyzer in zero-span mode, showing a FWHM of 550(25) MHz. The dashed curve shows the calculated Doppler-broadened absorption coefficient of the pump light; the individual components due to the $F^\prime = 1$ and $F^\prime = 2$ excited hyperfine levels are shown as dotted curves. Without any fitting or free parameters, the generated signal data is proportional to the MB absorption distribution of the pump transitions. (c) The generated signal amplitude responds roughly linearly to increased pump optical power. (d) The response of the generated signal amplitude to microwave powers from SG1. The largest pump and idler powers measured correspond to intensities of about 3 \& 7 mW/cm$^2$. {\color{black}Error bars represent the standard deviation.}}
		\label{fig:experiment}
	\end{figure*}
	
	\begin{figure*}
		\includegraphics{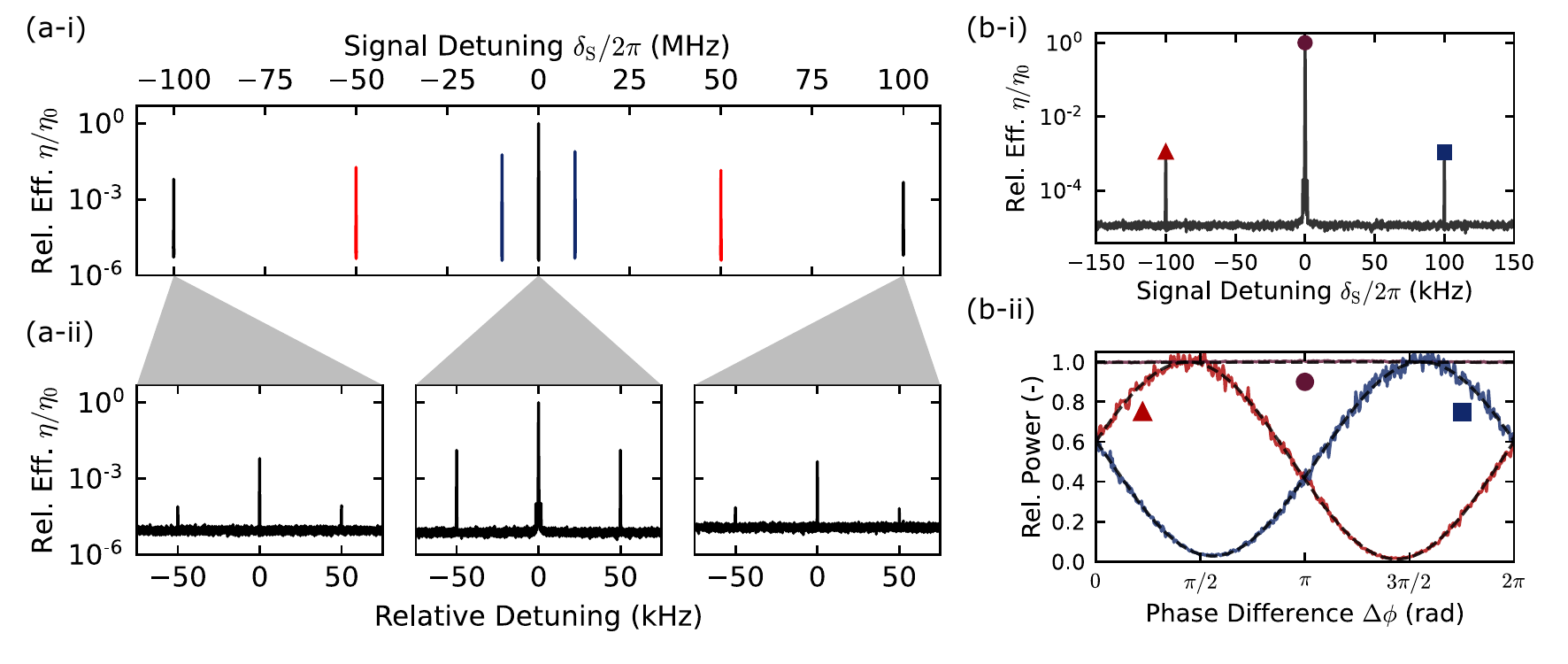}
		\caption{{\color{black} \bf{ Sideband generation and interference:}}(a-i) Multichannel generation using a fixed microwave amplitude modulation frequency of $\omega_{\rm AM} / 2\pi = 50~{\rm kHz}$ and a variable pump phase modulation frequency $\omega_{\rm PM} / 2\pi$ of 10 (blue), 50 (red) and 100 (black) MHz. The relative sideband amplitude decreases for increased $\omega_{\rm PM}$ because the pump light accesses velocity classes with decreased occupation. (a-ii) Zoomed-in data revealing AM sidebands flanking each of the central and SB PM peaks. (b) When the two modulation frequencies are equal, i.e. $\omega_{\rm PM} = \omega_{\rm AM}$, the PM and AM sidebands overlap and interfere. The relative phase difference  $\Delta \phi$ between the modulating waves controls the height of the {\color{black} correlated upper (\ding{110}, blue) and lower (\ding{115}, red) generated sidebands}. (b-i) The generated spectrum at zero relative phase. (b-ii) The relative power amplitude response of the peaks have a visibility of 97\%, while the central peak {\color{black}(\ding{108}, purple)} is independent of the relative phase. The two SB amplitudes differ from each other by $(\pi - 0.38)$ radians of phase. The dashed lines show independent fits to a cosine.}
		\label{fig:data}
	\end{figure*}
	
	\subsection{Experimental results}
	\indent First, we show that light is generated on the $|a\rangle \rightarrow |b\rangle$ transition. The $\chi^{(2)}$ interaction mediated by these atoms between the pump and microwave fields results in a sum-frequency process that generates a corresponding signal optical field. By combining the generated optical light with an optical local oscillator, we detect the generated light at the heterodyne beat frequency $f_{\rm het,~S}$ (see Methods \& {\color{black} Fig. 3(a)}).

	\indent To probe the frequency-tunable nature of the conversion process, we scan the pump laser frequency across the Doppler width. At any instant during the scan, the pump field interacts with one velocity class of atoms, which see this frequency on resonance with the pump transition. The generated signal reaches a maximum amplitude between the $F=2 \rightarrow F^\prime = 1$ and $F^\prime = 2$ transitions [Fig.~\ref{fig:experiment}(b)], with a full width at half maximum (FWHM) of 550(25) MHz. This generated spectrum corresponds to the combined Doppler-dependent absorption coefficients for the $F=1 \rightarrow F^\prime=1,~2$ transitions, which has a FWHM of 540 MHz. The $F^\prime=3$ state does not participate in this nonlinear process, since selection rules prohibit an electric-dipole transition to the $F=1$ ground state. Similarly, the pump transition from $F=2 \rightarrow F^\prime=0$ is electric-dipole forbidden. {\color{black} We also characterize the effects of various pump and idler powers as shown in Figs.~\ref{fig:experiment}(c-d). For a good signal-to-noise ratio, we typically operated with a pump power near $76~\mu\rm W$ and idler power of $3.2~\rm mW$.}
	
	\indent Next, we explore the multi-channel character of the conversion process. The pump laser is tuned halfway between the $F=2 \rightarrow F^\prime=2~\&~3$ transitions. By modulating either the idler or the pump, multiple frequency components are generated in the signal: First, we amplitude-modulate (AM) the microwave idler signal at $\omega_{\rm AM}$ and observe sidebands in the signal at $\omega_S~\pm~\omega_{\rm AM}$. The conversion bandwidth from idler modulation is limited to a maximum of 1 MHz by the width of the cavity resonance. Next, we additionally phase-modulate (PM) the pump using a fiber EOM in the optical path and observe that the idler and pump combine together as dictated by the sum-frequency process [Eq.~(\ref{eq:signal})]. The MB width and density determine the bandwidth and amplitudes of the pump-modulated generated sidebands. Both types of sideband generation are illustrated in Fig.~\ref{fig:data}(a), including simultaneous idler AM and pump PM. In the three cases depicted, $\omega_{\rm AM} / 2\pi = 50~{\rm kHz}$ is fixed and $\omega_{\rm PM} / 2\pi$ is varied from $10~\rm MHz$ to $100~\rm MHz$.
	
	\indent When the pump and idler are modulated at the same frequency, $\omega_{\rm PM} = \omega_{\rm AM}$, the sidebands from each input overlap and interfere {\color{black}[Fig.~\ref{fig:data}(b-i)]}. By tuning the relative phase of the modulating waves, one sideband or the other can be selectively enhanced or eliminated, with a relative visibility of 97\% [Fig.~\ref{fig:data}(b-ii)], in qualitative agreement with the calculated variation shown in Fig.~\ref{fig:theory}(d).
	
	The efficiency of conversion from microwave to signal photons is defined as the rate of signal photon production to the rate of microwave photons residing in the cavity mode~\cite{Adwaith2019, Han2018}
	\begin{equation}
		\eta \equiv \frac{P_{\rm signal} / \hbar \omega_{\rm signal}}{P_{\rm idler} / \hbar \omega_{\rm idler}}.
		\label{eq:effciency}
	\end{equation}
	Further details on this calculation are included in {\color{black} Supplementary Notes 2, 3 and 6}. We determine a maximum conversion efficiency of {\color{black}$\eta_0 = 1.5(1) \times 10^{-9}$}. Efficiency in our system is primarily limited by a very low optical depth of 0.05(1). In comparison, cold atom conversion experiments have employed optical depths of up to 15~\cite{Han2018} and 120~\cite{Tu2022}. Our optical depth corresponds to a number density of about $4 \times 10^{16}~{\rm cm^{-3}}$ with $8.7(1) \times 10^8$ atoms in the interaction region. This atom number is $\sim10^9$ times lower than the $3.2(1) \times 10^{18}$ available microwave photons in the same volume. At room temperature, the thermal noise is at the level of 900 photons, which is indiscernible compared to the signals used here. We note that it is this noise, determined by the temperature of the microwave electronics and transmission components (rather than of the conversion medium), that sets the limit of low-photon-number operation.  Moving toward true quantum transduction requires reducing the noise in these components by operating at cryogenic temperatures~\cite{Kumar2023}.

	\section{Discussion}
	\indent The frequency tunability of the generated signal field results from the existence of a large Doppler width at room temperature. While this work uses a warm vapor cell of enriched $^{87}$Rb without an anti-relaxation coating or a buffer gas, a buffer gas would provide additional tunability due to collisional broadening~\cite{Peng2022, Zameroski2011, Ha2020}. This tunability is ultimately limited to the order of the ground-state hyperfine splitting of the atoms used.
	
	\indent While other conversion schemes display degrees of tunability of the output frequency~\cite{Zhu2020, Shen2022, Han2018, Tu2022}, these are often restricted by natural atomic or optical cavity resonance frequencies. In the warm-atom system, a continuous tunability over the entire Doppler width is possible. This frequency multiplexing capability enables multiple microwave sources to be interfaced with a single atomic transducer, and the conversion can be used to place differing sources in distinct optical frequency domains. Switching between different optical frequencies for a single source is also possible. Our scheme features a conversion bandwidth of 910(20) kHz [see Fig.~\ref{fig:characterization}(b)], which is primarily limited by the cavity linewidth. As previously demonstrated in a different context~\cite{Godone1999}, the reverse optical-to-microwave conversion process is achievable in an atomic three-wave mixing system.
	
	The high-contrast amplitude control of signal sidebands enabled by simultaneous pump and idler modulation [Fig.~\ref{fig:data}(b)] is equivalent to a tunable beam-splitter operating in the frequency domain~\cite{Lu2018a}. The modulation indices and the relative phase act together as reflection and transmission coeffiecients. For particular values of these coefficients we can selectively generate a chosen sideband channel in the optical domain while completely suppressing the other, resulting in single-sideband conversion.
	
	Thus, our neutral atom system is capable of performing as a “quantum frequency processor”~\cite{Lukens2017, Lukens2019, Olislager2010} with particularly low input-field intensities. This allows us not only to encode and manipulate information in multiplexed frequency bins but also to convert it from microwave to optical frequency domains.
	
	The measured conversion efficiency is comparable with magnon-based microwave-to-optical conversion schemes~\cite{Zhu2020, Shen2022}. The warm-atom efficiency is inherently limited by the intrinsic weakness of the magnetic-dipole microwave transition, as well as the ground-state spin decoherence $\gamma_{\rm I}$ which includes wall-collisions, transit-time decoherence, and Rb-Rb spin-exchange collisions (see {\color{black} Supplementary Note 5}). A larger optical beam combined with a buffer gas~\cite{Franzen1959, Finkelstein2023} and anti-relaxation coating on the cell walls~\cite{Li2017, Balabas2010, Chi2020, Jiang2009} would help reduce spin decoherence from the dark state and boost conversion efficiency. For example, we expect that a high-quality anti-relaxation coating and a modest 5 torr of neon would significantly reduce transit-time and wall effects, and reduce $\gamma_{\rm I} / 2\pi$ by a factor of $10^3$ to 100 Hz, limited by Rb-Rb spin-exchange. Everything else being equal, this would provide a $10^6$-fold improvement to the generation efficiency. Furthermore, based on \cite{Adwaith2019}, we expect that increased pump powers will produce a moderate 4-fold efficiency gain before the efficiency saturates due to absorption and incoherent scattering of the pump.
	
	In this proof-of-concept demonstration, we have shown that inhomogeneous Doppler broadening in a  warm vapor cell provides significant tunability of the output optical frequency in a microwave-optical transduction, resulting in frequency division multiplexing. The inhomogeneous width also enables simultaneous frequency up-conversion of multi-channel microwave inputs. Amplitude control of up-converted optical channels demonstrates analogous frequency domain beam splitting action across five orders in frequency. The frequency division multiplexing capability combined with amplitude control can make neutral atoms as quantum processors for information encoded in frequency bin qubits.
	
	\section{Methods}
	\begin{figure}[htb]
		
		\includegraphics{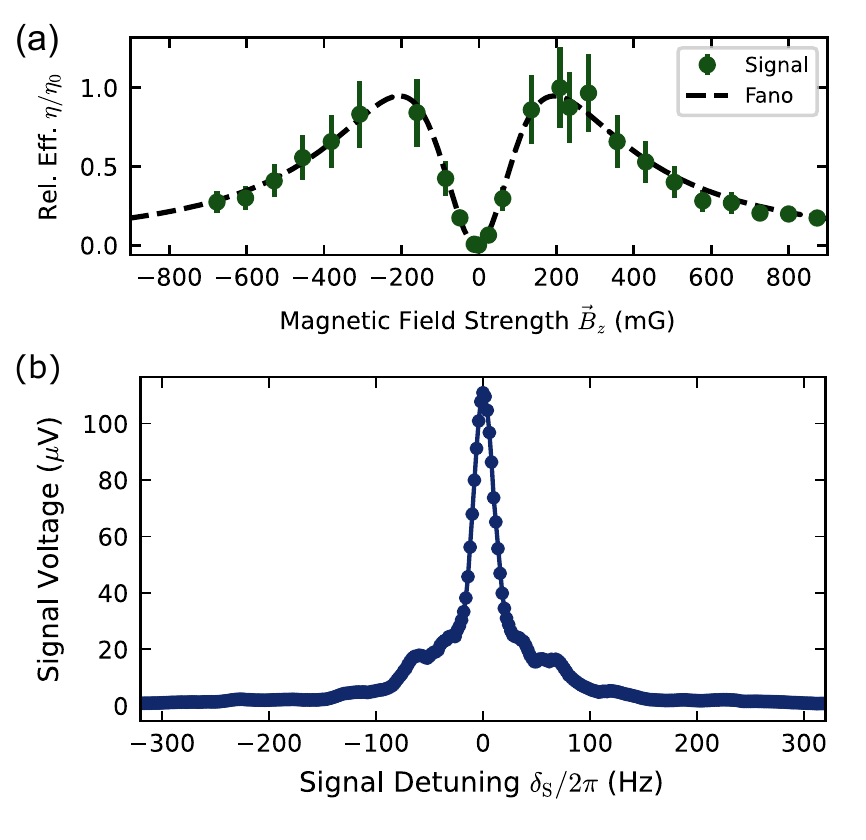}
		\caption{{\color{black} \bf{Magnetic field effect and generated linewidth:}} (a) The generated signal's response to a variable magnetic field strength $\vec{B}_z$ along the z-axis. We operate at 209 mG for maximum generation. We fit a symmetric Fano profile [Eq.~(\ref{eq:fano})] to these data, with fit parameters $\eta_0^\prime = 0.95(2)$, $\gamma_B = 200(4)~{\rm mG}$ and $B_0 = -5(3)~{\rm mG}$. This behavior might indicate the interaction and interference of multiple Zeeman level resonances within the decoherence width of our atoms. {\color{black}Error bars represent the standard deviation.} (b) A close up of the unmodulated generated line revealing a FWHM of 26(2) Hz.}
		\label{fig:response}
	\end{figure}
	
	\begin{figure}[htb]
		\includegraphics{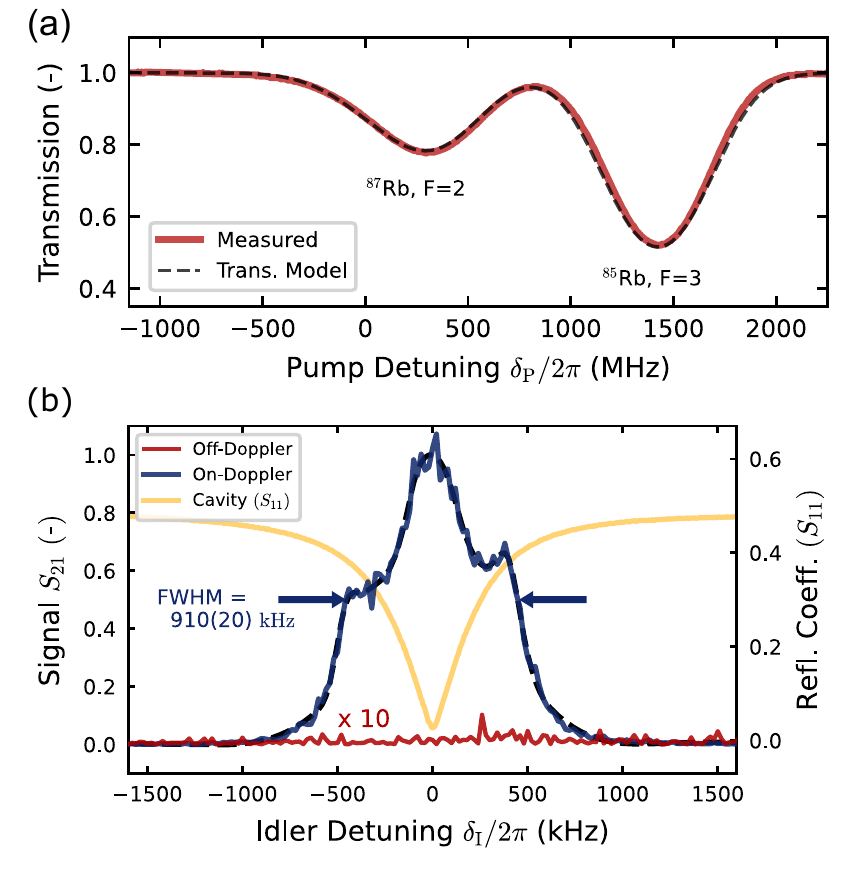}
		\caption{{\color{black} \bf{Frequency sweep and conversion bandwidth
					:}} (a) The Doppler-broadened transmission spectrum of a reference rubidium vapor cell (outside the microwave cavity) at room temperature, measured while sweeping the pump laser frequency. The dotted line shows a transmission model curve with good agreement to the measured trace. This trace was taken simultaneous to the data shown in Fig.~\ref{fig:experiment}(b); the positions of the two Doppler peaks are used to calibrate the horizontal frequency scale. (b) The raw generated signal during a scan of $\delta_I$, as measured with a vector network analyzer {\color{black} (VNA)} in S21 mode. We measure a conversion bandwidth (FWHM) of 910(20) kHz when the laser is locked on-resonance and with a $z$-magnetic field of 209 mG. The dashed line is a smoothing spline used to determine the bandwidth. No generation is visible when the laser is off-resonance. The cavity resonance has a FWHM of 417 kHz, as measured with the VNA in S11 mode.}
		\label{fig:characterization}
	\end{figure}

	\subsection{Experimental Setup}
	\indent The experimental setup can be seen in Fig.~\ref{fig:experiment}(a). The pump optical field is derived from an continuous-wave external cavity diode laser (MOGLabs CEL) operating at 780.24 nm. This laser is frequency-stabilized to the saturated absorption spectrum of a separate Rb reference vapor cell. We have the option to lock the laser frequency 80 MHz below the $F^\prime$=2-3 crossover transition with a locked linewidth of $\sim100~\rm kHz$. Alternatively, we can unlock and move the laser frequency off-resonance. The pump beam is then shifted upwards in frequency by $f_{\rm AOM} = 76.60~\rm MHz$ with an acousto-optical modulator and transmitted to the cavity-vapor cell apparatus via optical fiber. After the optical fiber, the polarization is filtered with a polarizing beam splitter (PBS) and the power of the pump light is stabilized with a servo controller (NuFocus LB1005) to $<1$\% fluctuations. The Rabi frequency of the pump laser $\Omega_{\text{P}} / 2\pi$ has a value of 3.5 MHz~\cite{Sibalic2017} and it is linearly polarized along the $z$-axis. 
	
	\indent Inside the cavity~\cite{Tretiakov2020, Ruether2021}, the idler microwave field is polarized along the $y$-direction. Orthogonal Helmholtz coils null the external magnetic field along the $x$ and $y$-axes and apply a weak DC magnetic field $\vec{B}_z$ along the $z$-axis. The magnetic field strength makes a striking impact on the conversion efficiency. As shown in Fig.~\ref{fig:response}(a), conversion is strongly suppressed at zero magnetic field and reaches maxima at $\sim\pm 200~\rm mG$. This behavior is well-described by a symmetric, generalized Fano lineshape~\cite{Caselli2018} that has the simplified form
	\begin{equation}
		\label{eq:fano}
		\eta(B_z) =  \eta_0^\prime \left[\frac{2 \gamma_B (B_z - B_0)}{\gamma_B^2 + (B_z - B_0)^2}\right]^2.
	\end{equation}
	
	\noindent This points to the participation and interaction of multiple Zeeman levels in the generation process. At Zeeman degeneracy, generation pathways may destructively interfere. The value $\gamma_B \times \gamma_0 = 140~{\rm kHz} \sim \gamma_I$, the ground state decoherence rate, where $\gamma_0$ is the ground state gyromagnetic ratio of $^{87}$Rb~\cite{SteckRb87}. Further discussion goes beyond the scope of this work and will be the topic of future investigation.
	
	\indent The generated signal light emerges from the cavity polarized along the $y$-axis, orthogonal to the pump~\cite{Campbell2022}. Before recoupling to an optical fiber, a half-waveplate is manually adjusted to optimize the signals at $f_{\rm het,~P}$ and $f_{\rm het,~S}$ for each respective measurement. The light is then combined in a 50:50 fiber {\color{black}BS} with local oscillator laser light that was not shifted by the acousto-optic modulator (AOM). The interference produces heterodyne beat notes at $f_{\rm het,~P} = f_{\rm AOM}$ and $f_{\rm het,~S} = \omega_\mu / 2\pi + f_{\rm AOM} = 6.911~282~\rm GHz$. A high bandwidth amplified photodiode (PD1, EOT ET-4000AF) detects the beating signals. This electronic signal is amplified (AMP, Mini-Circuits ZJL-7G+) and its amplitude is measured by a spectrum analyzer (SA, R\&S FSV3013). This heterodyne peak at $f_{\rm het,~S}$ has a FWHM of 26(2) Hz [see Fig.~\ref{fig:response}(b)], larger than the $< 2~\rm Hz$ linewidth of our microwave source, SG1. We presume the presence of additional perturbations (e.g. 60 Hz line noise) broadens and modulates our generated signal. 
	
	\subsection{Broad-tunability}
	\indent To access all the frequencies within the Doppler profile, the pump frequency is scanned across the Doppler profile with a laser frequency sweep. We trigger the SA off the sweep ramp and measure the heterodyne signal in zero-span mode with a resolution bandwidth of 2 kHz. Simultaneous to the sweep, we also trigger and record the transmission of pump light through a reference vapor cell of rubidium (natural abundance) placed near the microwave cavity and at ambient $21.6(3)~^{\circ}\rm C$. We find good agreement between the measured spectrum and a theoretical transmission model~\cite{Siddons2008}. Using the $^{87}$Rb and $^{85}$Rb transmission dip extrema, we calibrate our sweep rate to 17.2 MHz/ms and convert the time axis of the series to frequency [Fig.~\ref{fig:characterization}(a)]. 
	
	\subsection{Multichannel Conversion}
	\indent We generate a multichannel optical signal field in two different ways: first, the signal generator (SG1) supplying the idler field is amplitude-modulated at a variable $\omega_{\rm AM}$. We note that due the finite microwave cavity resonance (see {\color{black} Supplementary Note 1}), $\omega_{\rm AM} / 2\pi$ can be tuned up to ~455(10) kHz before the sidebands become significantly suppressed. Second, we modulate the pump light by using a fiber phase-modulator EOM (EOSPACE PM-0S5-20-PFA-PFA-780) inserted into the pump light optical path. We drive the EOM with a second signal generator (SG2, SRS SG384) at a variable $\omega_{\rm PM}$. This produces sidebands in the pump optical spectrum, and corresponding sidebands are visible in the generated spectrum around the heterodyne frequency $f_{\rm het,~S}$. The SA and SG2 are both phase-locked to the internal 10 MHz clock reference of SG1. In this second case, $\omega_{\rm PM}$ can be tuned over the Doppler FWHM before the sidebands become significantly suppressed.
	
	\subsection{Conversion Bandwidth}
	We characterize the conversion bandwidth of our microwave-to-optical conversion process. To do this, we replace SG1 and SA with Ports 1 and 2, respectively of a VNA (Keysight E5063A). The generated signal is measured on the VNA in S21 mode and is shown in Fig.~\ref{fig:characterization}(b). When the laser is locked on resonance and at the optimal magnetic field of 209 mG, a clear peak in the S21 signal is visible with FWHM = 910(20) kHz. This width is compared to the 417 kHz width of the cavity resonance. When the laser is off-resonance, no generation is apparent. 

	\section{Data Availability}
	{\color{black} The datasets generated during and/or analysed during the current study are available in the University of Alberta Dataverse Borealis repository, \url{https://borealisdata.ca/dataset.xhtml?persistentId=doi:10.5683/SP3/LGCNHZ}.}

	\begin{acknowledgments}
		\indent We gratefully acknowledge that this work was performed on Treaty 6 territory, and as researchers at the University of Alberta, we respect the histories, languages, and cultures of First Nations, Métis, Inuit, and all First Peoples of Canada, whose presence continues to enrich our vibrant community
		
		\indent We are indebted to Clinton Potts for machining the microwave cavity used in this work. We are also grateful to the John Davis, Mark Freeman, Robert Wolkow, and the Vadim Kravchinsky groups for generously lending equipment used in these experiments, and to Ray DeCorby for helpful feedback on this work.
		
		\indent \color{black}This work was supported by the University of Alberta; the Natural Sciences and Engineering Research Council, Canada (Grants No. RGPIN-2021-02884 and No. CREATE-495446-17); the Alberta Quantum Major Innovation Fund; Alberta Innovates; the Canada Foundation for Innovation; and the Canada Research Chairs (CRC) Program.
	\end{acknowledgments}
	
	\section{Author Contributions}
	{\color{black}BDS and AN developed the concept for the experiment, and AN provided the models used. Experiments were performed by BDS, BB, and AN and analyzed by BDS with input from AN. The manuscript was initially prepared by BDS and AN, and all authors contributed to review and editing. LJL supervised the project.
	}
	
	\section{Competing interests}
	{\color{black} The authors declare no competing interests.}
	

	\onecolumngrid
	\appendix
	\section{Supplementary Information}
	\subsection{Supplementary Note 1: Microwave Cavity \& Vapor Cell}
	\noindent The cylindrical microwave cavity is machined out of low-oxygen copper, with an internal diameter of 58 mm and length of 53(2) mm. The resonance frequency of the cavity TE011 mode is precisely tuned to the rubidium ground state transition, $\omega_0 = 2\pi \times 6.834~682~\rm GHz$, by a micrometer attached to the cavity endcap. Microwave power to the cavity is supplied by a low-noise signal generator (SG1, AnaPico APSIN12G). As the ground state magnetic-dipole transition is weak, the cavity mode enhances this transition probability. Inside the cavity is a rectangular quartz cell with external dimensions of 30 x 12.5 x 12.5 mm (1 mm thick walls) containing enriched $^{87}$Rb. It has a cylindrical stem that protrudes from the {\color{black} cavity face opposite the endcap} and is gently clamped in place by a support structure water-cooled to 16$^\circ$C. In addition, the cavity is wrapped with silicone thermal heating pads and aluminum foil, and is temperature stabilized at 37$^\circ$C. We discovered that excess rubidium inside the main volume deposits and forms a thin metallic film on the inner cell walls. This metalization scatters the internal microwave mode, thereby lowering the internal cavity quality factor. The cooled stem pulls the excess rubidium out of the cavity, and the heated cavity vaporizes the deposited metal. With both heating and cooling, we measure the loaded quality factor (Q) of our cavity with a vector network analyzer in S11 mode to be $Q = 9000 +/- 100$. The rubidium vapor density is ultimately set by the temperature of the cold stem through a ``reservoir'' effect~\cite{Karaulanov2009}. From our measured pump optical depth of 0.03(1) we calculate a number density of $1.7(6) \times 10^{16}~\rm m^{-3}$. \\

	\subsection{Supplementary Note 2: Cavity Mode Volume}
	
	\noindent We performed COMSOL simulations of our microwave cavity and the $\rm TE_{011}$ mode of interest. With the simulation results, we calculated the {\color{black} magnetic} mode volume from the equation~\cite{Lambert2020}: 
	
	\begin{equation}
		\label{S1}\tag{S1}
		V_{\rm m} = \frac{\int_{V_{\rm cav}}|\mathbf{B}(\mathbf{r})|^2 / \mu(\mathbf{r})}{\left(|\mathbf{B}(\mathbf{r})|^2 / \mu(\mathbf{r})\right)_{\rm max}}  
	\end{equation}
	to be $9.19~\rm cm^3$, where $V_{\rm cav}$ is the internal volume of the microwave cavity.

	\subsection{Supplementary Note 3: Interaction Volume}
	\noindent We determine the conversion interaction volume to be the region inside the vapor cell illuminated by the pump laser beam. We make the assumption that the microwave cavity mode is uniform over this volume. The pump beam incident to the microwave cavity is collimated with a $1/e^2$ radius of $r_0 = 0.87~\rm mm$ and a power of $75~\rm uW$. Due to the hole aperture of $r_{\rm h} = 1.5~{\rm mm}$ radius, about 95\% of the incident light passes through. The effective uniform beam has a area of: 
	\begin{align}
		A_{\rm eff} = \int_0^{r_{\rm h}} r e^{-(r/r_0)^2}dr &= \pi r_0^2 \left(1 - e^{-(r_{\rm h}/r_0)^2}\right) \label{S2}\tag{S2} \\
		&\simeq 0.95 \times \pi r_0^2 \label{S3}\tag{S3}\\
		&= 2.21(4)~{\rm mm^2} \label{S4}\tag{S4}
	\end{align}
	and the volume overlapping with the atoms is $V_{\rm int} = 19(1)~{\rm mm^3}$.

	\subsection{Supplementary Note 4: Numerical modeling for generated field amplitude}
	
	Our numerical modeling follows a formalism developed for a 
	closed three level atomic system interacting with idler, pump and signal fields~\cite{Adwaith2021}. 
	
	The Hamiltonian of a generic three level atomic system interacting with signal ($\omega_{\text{S}}$), idler ($\omega_{\text{I}}$) and pump ($\omega_{\text{P}}$) fields is considered and the non-resonant time varying terms in the Hamiltonian removed under a rotating-wave approximation. 
	The detunings of the signal, idler and pump fields from their respective transitions (see Fig. 1(c) in the paper) are denoted as $\delta_{\text{S}}$, $\delta_{\text{P}}$ and $\delta_{\text{I}}$. 
	An  unitary transformation of the form 
	\begin{align}
		\hat{{U}}(t)
		= \vert b\rangle \langle b\vert + e^{i((\omega_{\text{S}}-\omega_{\text{P}})t-(k_{\text{S}}-k_{\text{P}})z)}\vert c\rangle \langle c\vert + e^{i(\omega_{\text{S}}t-k_{\text{S}}z+\Delta \phi)}\vert a\rangle \langle a\vert
		\label{S5}\tag{S5}
	\end{align}
	is applied
	to obtain a time-independent Hamiltonian. The resultant Hamiltonian $\hat{H}(\bm{r},\bm{v})$ under a three-photon resonance condition ($\delta_{\text{S}} - \delta_{\text{P}} + \delta_{\text{I}}$ = 0) is  
	\begin{align}
		\hat{H}(\bm{r},\bm{v}) =& -\Delta_\text{S} (\bm{v}) \vert a\rangle \langle a\vert -\left(\Delta_\text{S}(\bm{v}) - \Delta_\text{P}(\bm{v})\right)\vert c\rangle \langle c\vert -\frac{\Omega_\text{S}(\bm{r})}{2}\vert a\rangle \langle b\vert \nonumber \\
		& - \frac{\Omega_\text{P}(\bm{r})}{2}\vert a\rangle
		\langle c\vert -\frac{\Omega_{\text{I}}(\bm{r})e^{-i\Delta\Phi}}{2}\vert c\rangle \langle b\vert + \text{H.C.},
		\label{S6}\tag{S6}
	\end{align}
	where H.C. denotes the Hermitian conjugate and $\Delta\Phi$ the relative phase between all the three fields. The zero-energy reference for the Hamiltonian is taken to be the level $\vert b \rangle$ (see Fig. 1(c) in the paper). The Doppler-shifted detunings of the signal and pump field, as
	seen by an atom moving with velocity $\bm{v}$, are given by $\Delta_\text{S} (\bm{v}):=\delta_\text{S}- \bm{k}_\text{S}\cdot\bm{v}$ and $\Delta_\text{P} (\bm{v}):=\delta_\text{P}- \bm{k}_\text{P}\cdot\bm{v}$. Here $\Omega_\text{S} (\bm{r}) ={\bm{d}_{ba}\cdot\bm{E}_{\text{S}} (\bm{r})}$, $\Omega_\text{P} (\bm{r}) ={\bm{d}_{ca}\cdot\bm{E}_{\text{P}} (\bm{r})}$ and $\Omega_{\text{I}}(\bm{r}) ={\boldsymbol{\mu_{bc}}\cdot\bm{B}_{\text{I}}(\bm{r})}$ are the Rabi frequencies of the signal, pump and idler fields; and $\bm{d}_{\text{ij}}$ and $\boldsymbol{\mu}_{\text{ij}}$ are the electric and magnetic dipole matrix elements, respectively. All optical fields are taken to be propagating along the $\bm{z}$ direction, with $\bm{E}_{\text{S}}$ and $\bm{E}_{\text{P}}$ denoting the electric fields of the signal and pump fields, respectively, and $\bm{B}_{\text{I}}$ is the microwave magnetic-field flux density.

	The populations and coherences of the three-level system described by the Hamiltonian are denoted by the density matrix elements ${\hat{\rho}_{ij}}$ with $i, j \in (a, b, c)$. The well known dynamical evolution equation of the density matrix elements in a rotated frame  $\hat{\sigma}(t) = \hat{{U}}(t)\hat{\rho}(t)\hat{{U}}^{\dagger}(t)$ is:
	\begin{equation}
		\frac{\partial {\hat{\sigma}}(\bm{v},z,t)}{\partial t}=-i[\hat{H}(\bm{v},z),\hat{\sigma}(\bm{v},z,t)]+\sum_{k=1}^3\mathcal{L}(\hat{a}_k)\hat{\sigma}(\bm{v},z,t) 
		\label{S7}\tag{S7}
	\end{equation}
	with $\mathcal{L}(\hat{a}_k)$ the Lindblad superoperator
	\begin{equation}
		\mathcal{L}(\hat{a})\hat{\sigma}(\bm{v},z,t)=\hat{a}\hat{\sigma}(\bm{v},z,t)\hat{a}^\dagger-\frac{1}{2} 
		\left\{\hat{\sigma}(\bm{v},z,t),\hat{a}^\dagger\hat{a}\right\}
		\label{S8}\tag{S8}
	\end{equation}
	acting on operators\hspace{1mm} $\hat{a}_1 = \sqrt{\gamma_{\text{I}}}\vert b\rangle \langle c\vert$,\hspace{1mm} ,\hspace{1mm} $\hat{a}_2 = \sqrt{\gamma_{\text{S}}}  \vert b\rangle \langle a\vert$,\hspace{1mm} and $\hat{a}_3 = \sqrt{\gamma_{\text{P}}} \vert c\rangle \langle a\vert$.
	with~$\gamma_{\text{I}}$ and $\gamma_{\text{P}}=\gamma_{\text{S}}$ representing natural linewidths of levels~$\vert c \rangle$ and $\vert a \rangle $ respectively.

	In the limit of weak signal and idler fields, $\Omega_{\text{S}}<\Omega_{\text{I}}\ll\Omega_{\text{P}}$
	we assume that all the population is in ground state i.e., $\sigma_{bb} \approx{1}$.  With this approximation, we solve Eq.~(\ref{S6}) under steady state condition to obtain an expression for $\sigma_{ab}$:
	\begin{align}
		\sigma_{ab}(\bm{v},z) = \frac{ {2i\Gamma_2(v)\Gamma_3(v)\Omega_{\text{S}}(\bm{z})} + {\Gamma_3(v)\Omega_{\text{P}}(\bm{z})\Omega_{\text{I}}(\bm{z})e^{-i\Delta\Phi}}}{4\Gamma_1(v)\Gamma_2(v)\Gamma_3(v) + \Gamma_3(v)|\Omega_{\text{P}}(\bm{z})|^2 + \Gamma_2(v)|\Omega_{\text{I}}(\bm{z})|^2}  ,
		\label{S9}\tag{S9}
	\end{align}
	where
	\begin{align}
		\Gamma_1(v)=&\frac{\gamma_{\text{S}}}{2}+\frac{\gamma_{\text{P}}}{2}-i \Delta_{\text{S}}(v),\nonumber\\
		\Gamma_2(v)=&\frac{\gamma_{\text{I}}}{2}- i (\Delta_{\text{S}}(v) - \Delta_{\text{P}}(v)), \tag{S10} \\
		\Gamma_3(v)=&\frac{\gamma_{\text{S}}}{2}+\frac{\gamma_{\text{P}}}{2}+\frac{\gamma_{\text{I}}}{2}- i\Delta_{\text{P}}(v).\nonumber
		\label{S10}
	\end{align}
	
	In Eq.~(\ref{S9}) the coefficient of $\Omega_{\text{S}}$ in the first numerator term is proportional to the linear susceptibility of signal field if  a seed optical signal field is also present in the experiment. The coefficient of $\Omega_{\text{P}}\Omega_{\text{I}}$ in the second numerator term is proportional to the hybrid second order susceptibility of signal field. This nonlinear second-order susceptibility results from a combined magnetic and an electric dipole transition induced by the microwave idler and optical pump field respectively resulting in a {\it generated} optical signal field.
	
	Using the  slowly-varying envelope approximation we calculate the propagation equation for the signal field inside our atomic nonlinear medium as:
	\begin{equation}
		\frac{\partial \Omega_\text{S}(v)}{\partial z}=i \eta_{\text{S}}(v) \sigma_{ab}(v,z),
		\label{S11}\tag{S11}
	\end{equation}
	where $\eta_{\text{S}}(v)=({\bm{d}_{\text{ab}}^2\omega_\textsubscript{S}N(v)}/{ 2 \epsilon_{0}c)}$ is the coupling constant for the signal field, and $N(v)$ is the number density of atoms in a velocity class $v$. $N(v)$ is taken from the Maxwell-Boltzmann distribution of velocities of atoms at 300 K. The effect of spatial variation of microwave intensity inside the cavity is ignored due to the smallness of the interaction region within which the atoms and all the three fields interact. Over this small volume the microwave intensity is assumed to be constant. We therefore maintain $\Omega_{\text{I}}(\bm{z})\equiv\Omega_{\text{I}}(0)$ in our theoretical analysis for all velocity class of atoms. In addition, we work under the undepleted pump approximation for our pump field. This fact is incorporated by taking $\Omega_\text{P} (\bm{z})\equiv\Omega_{\text{P}}(0)$. We thus solve Eq.~(\ref{S11}) with spatially uniform pump and idler fields. The resulting solution for the signal field for an atomic sample of length $l$ is:
	\begin{equation}
		\Omega_{\text{S}}(v,{l}) = \Omega_{\text{S}}(0) e^{-\beta(v) {l}} + \frac{i \Omega_{\text{P}}(0) \Omega_{\text{I}}(0)e^{-i\Delta\Phi}}{2\Gamma_2(v)} \left(e^{-\beta(v) {l}} - 1\right)
		\label{S12}\tag{S12}
	\end{equation}
	where 
	\begin{align}
		\beta(\nu) =\frac{{2\eta_{\text{S}}(v)\Gamma_2(v)\Gamma_3(v)}}{{4\Gamma_1(v)\Gamma_2(v)\Gamma_3(v) + \Gamma_3(v)|\Omega_{\text{P}}(0)|^2 + \Gamma_2(v)|\Omega_{\text{I}}(0)|^2}}
		\label{S13}\tag{S13}
	\end{align}
	It is  noted that $\beta(\nu)$ is, in general, a complex number.
	
	The signal field amplitude at the end of an atomic medium of length $l$ is given by Eq.~(\ref{S12}). The first term in Eq.~(\ref{S12}) represents phase changes and losses suffered by an {\it input} signal field due to linear susceptibility  of the atomic field at the frequency of the signal field. In the current experiment there is no {\it input} signal field. Therefore $\Omega_{\text{S}}(0) = 0$.
	Thus the solution in Eq.~(\ref{S12}) reduces to 
	\begin{equation}
		\Omega_{\text{S}}(v,{l}) =  \frac{i \Omega_{\text{P}}(0) \Omega_{\text{I}}(0)e^{-i\Delta\Phi}}{2\Gamma_2(v)} \left(e^{-\beta(v) {l}} - 1\right)
		\label{S14}\tag{S14}
	\end{equation}
	Eq.~(\ref{S14}) represents the {\it generated} signal field amplitude for each velocity class $v$ of atoms through a second-order nonlinear process which combines the idler field $\Omega_{\text{I}}(0)$ and the pump field $\Omega_{\text{P}}(0)$.

	\subsection{Supplementary Note 5: Spin Decoherence Estimates}
	In the main manuscript, we group all spin decoherence effects -- including transit-time decoherence $\Gamma_T$, spin-exchange collision $\Gamma_{\rm SE}$, and wall depolarization $\Gamma_{\rm wall}$ -- into a single decoherence rate $\gamma_{\rm I} = \Gamma_T + \Gamma_{\rm SE} + \Gamma_{\rm wall}$, which we estimate from double-resonance measurements to be $\gamma_{\rm I} / 2 \pi = 130(30)~\rm kHz$. As seen in the denominator of Eq.(1), the strength of the generated signal is suppressed by a factor of $\gamma_{\rm I}$, implying that greater efficiency is possible by reducing decoherence.
	
	The individual decoherence components are estimated for room temperature as follows: the transit-time decoherence rate is given by $\Gamma_T / 2 \pi \sim \bar{v} / 4\pi r_{\rm eff}$~\cite{Budker2008}, the average thermal velocity $\bar{v} \simeq 430~\rm m/s$ divided by the effective beam radius, and has a value of about $\Gamma_T / 2 \pi \sim 20~\rm kHz$. Next, the spin-exchange collision rate is $\Gamma_{\rm SE} / 2 \pi = \bar{v}n\sigma_{\rm SE}$, where $n$ is the atomic density and $\sigma_{\rm SE} = 1.9\times 10^{-14}~\rm cm^2$ is the spin-exchange cross-section between two rubidium atoms~\cite{Happer2010}, giving $\Gamma_{\rm SE} / 2 \pi \sim 100~\rm Hz$. This value is relatively small, implying that the atoms have a mean free path of $\lambda = 1 / n\sigma_{\rm SE} \sim 2~\rm m$ before experiencing a spin-flip collision with another atom. This distance is much larger than any cell dimension, implying that the atoms are mostly ballistic and wall collisions dominate decoherence.
	
	Precisely calculating the effect of wall collisions is an involved process~\cite{Happer2010, Happer1972}, but an upper-bound estimate is possible. We define a characteristic cell length $\ell$ as the ratio of the cell volume to the surface area~\cite{Happer2010}: 
	
	\begin{equation}
		\label{S15}\tag{S15}
		\ell = V / S \simeq \rm 3~cm^3 / 14~cm^2 \simeq 0.2~cm.
	\end{equation}
	
	This represents the average distance between wall collisions. The average wall collision rate is then simply $\Gamma_{\rm wall} / 2 \pi \lesssim \bar{v} / \ell \simeq 120~\rm kHz$. Again, this upper bound does not consider adsorption binding energies or dwell times~\cite{Happer1972}.
	
	Consider, for example, we added 5 torr of neon buffer gas and a high-quality, alkene anti-relaxation wall coating to our vapor cell. In this case, the rubidium atoms in the buffer gas have a mean free path of $\sim 100~\rm \mu m$, putting this in a diffusive regime. Wall effects are effectively eliminated by the coating and the slower atomic diffusion. The spin-exchange and relaxation cross sections between the rubidium and the neon are much smaller, $\sigma_{\rm SE}^\prime \sim 10^{-23}~\rm cm^2$~\cite{Walker1989}, and will not contribute significantly. However, it will be impossible to eliminate the rubidium-rubidium spin-exchange collisions~\cite{Budker2008}, so this factor will dominate $\gamma_{\rm I} / 2\pi$ with an ultimate value of $\sim 100~\rm Hz$.

	\subsection{Supplementary Note 6: Conversion Efficiency}:
	
	\noindent The generated optical power is weaker than the residual pump by a factor determined by the relative amplitudes of the two heterodyne peaks $P_{\rm SA,~pump}$ and $P_{\rm SA, signal}$ measured (in dBm) on the spectrum analyzer:
	\begin{equation}
		\label{S16}\tag{S16}
		P_{\rm signal} / P_{\rm pump} = 10^{\left(P_{\rm SA,~signal} - P_{\rm SA, pump} + \zeta \right) / 10}.
	\end{equation}
	The correction term $\zeta$ is the combined frequency response and insertion loss difference between the two heterodyne frequencies. It is calibrated from the response of a beating signal between the local oscillator light and a weak beam from a separate tunable laser.
	
	To estimate the available microwave photon rate, we first determine the fraction of the source power $P_{\rm cav} / P_0$ which is coupled into the cavity. From an S21 measurement of the cable, its insertion loss $IL$ was determined from S21 to be -2.928(3) dB and the cavity absorbs $\Delta RL=-23~\rm dB$ beyond this, meaning that all but 0.5\% of the incident power is transmitted to the cavity.
	
	\begin{equation}
		\label{S17}\tag{S17}
		P_{\rm cav} / P_0 = 10^{(IL - \Delta RL)/10}
	\end{equation}
	{\color{black} Now $P_{\rm cav}$ is scaled by the ratio of the interaction to magnetic mode volumes. This gives the fraction of the microwave power that transitions the atoms between the two ground states.} Therefore, the microwave power involved in the conversion is:
	\begin{equation}
		\label{S18}\tag{S18}
		P_{\rm idler} = {\color{black}P_0}\frac{V_{\rm int}}{V_{\rm m}}10^{(IL - \Delta RL)/10}
	\end{equation}
	A critical difference between this efficiency calculation and the previous work of Adwaith, et al.~\cite{Adwaith2019} is that they divided by $V_{\rm cav}$ instead of the more accurate $V_{\rm m}$~\cite{Lambert2020}. 
	Finally, we calculate the efficiency from Eq.~(5) from the main text to be:
	\begin{align}
		\label{S19}\tag{S19}
		{\color{black}\eta = \frac{P_{\rm pump}}{P_0}} \frac{V_{\rm m}}{V_{\rm int}} \frac{\omega_I}{\omega_S} \frac{10^{\left(P_{\rm SA,~signal} - P_{\rm SA, pump} + \zeta \right) / 10}}{10^{(IL - \Delta RL)/10}}.
	\end{align}

	\subsection{Supplementary Note 7: Modulated Sideband Interference}
	A derivation of sideband amplitudes of the generated signal field with amplitude modulation of the idler field and phase modulation of pump field is given below. The derivation is for equal modulation frequencies $\delta$ with the relative phase of modulations  taken as $\Delta \phi$.\\
	\indent The pump field phase modulation is represented as 
	\begin{equation}
		\label{S20}\tag{S20}
		\Omega_{p0} \sin[(\omega_{\text{P}} t + a_p \sin(\delta t)].
	\end{equation}
	Here $\Omega_{p0}$ is the initial value of the pump field amplitude and $a_p$ is the phase modulation index. The idler field is amplitude modulated as 
	\begin{equation}
		\label{S21}\tag{S21}
		\Omega_{I0} [1+a_m \sin(\delta t + \Delta \phi)]\sin(\omega_{\text{I}} t),
	\end{equation}
	with amplitude modulation index $a_m$. The generated field is obtained by multiplying the idler and pump fields. This is equivalent to summing their frequencies in the frequency domain.
	The resultant expression for the generated field under small phase modulation index values is
	\begin{equation}
		\label{S22}\tag{S22}
		\Omega_{\text{s}}(t) = \Omega_{p0} \Omega_{I0}  \sin(\omega_{\text{I}} t) \left[1+a_m \sin(\delta t +\Delta \phi)\right] \left[\sin(\omega_{\text{P}} t) + a_p \cos(\omega_{\text{P}} t)\sin(\delta t) \right]
	\end{equation}
	Using trigonometric identities, Eq.~(\ref{S22}) can be grouped based on the generated frequencies in the signal field through the sum frequency process. The relevant terms are the central frequency at $\omega_s = \omega_I + \omega_P$,  the first positive sideband at $\omega_s + \delta$ and the first negative sideband at  $\omega_s - \delta$: 
	\begin{align}
		\label{S23}\tag{S23}
		&\Omega_{p0} \Omega_{I0} \cos(\omega_s t) \\
		\label{S24}\tag{S24}
		&\Omega_{p0} \Omega_{I0} A_{+} \sin\left([\omega_s + \delta] t + \theta_+\right)\\
		\label{S25}\tag{S25}
		&\Omega_{p0} \Omega_{I0} A_{-} \sin\left([\omega_s - \delta] t + \theta_- \right)
	\end{align}
	The amplitudes of the first positive and negative sidebands ($A_\pm$) are:
	\begin{align}
		\label{S26}\tag{S26}
		& A_{+} = \sqrt{a_m^2+ a_p^2 + 2 a_p a_m \sin(\Delta \phi) } \\
		\label{S27}\tag{S27}
		& A_{-} = \sqrt{a_m^2+ a_p^2 - 2 a_p a_m \sin(\Delta \phi) } 
	\end{align}
	The  phases of the first sideband oscillations ($\theta_{\pm}$) are: 
	\begin{align}
		\label{S28}\tag{S28}
		& \tan (\theta_+) = \frac{a_m \cos(\Delta \phi)}{ a_m \sin(\Delta \phi) + a_p} \\
		\label{S29}\tag{S29}
		& \tan (\theta_-) = \frac{ a_m \sin(\Delta \phi) - a_p}{a_m \cos(\Delta \phi)} 
	\end{align}

\end{document}